\documentstyle[sprocl]{article}
\def\Journal#1#2#3#4{{#1} {\bf #2}, #3 (#4)}
\def\PLA{{\em Phys. Lett.}  A}

\begin{document}

\title{SOLVING THE HORIZON PROBLEM WITH
A DELAYED BIG-BANG SINGULARITY}

\author{Marie-No\"elle C\'EL\'ERIER}

\address{D\'epartement d'Astrophysique Relativiste et de Cosmologie \\
Observatoire de Paris--Meudon \\
5 Place Jules Janssen 92195 Meudon C\'edex FRANCE }

\maketitle\abstracts{
The hot Big-Bang standard model for the evolution of the universe, despite strong successes, lets unresolved a number of problems. One of its main drawbacks, known as the horizon problem, was until now thought to be only solvable by an inflationary scenario. Here is proposed a class of inhomogeneous models of universe, getting rid of some of the worst drawbacks of standard cosmology. The horizon problem is solved by means of an initial singularity of ``delayed'' type and without need for any inflationary phase. The flatness and cosmological constant problems disappear.}
%\end{abstract}

%\medskip
%\noindent$\hrulefill $

%\bigskip

%\end{frontmatter}

	In this work, completed in collaboration with Jean Schneider \cite{cs}, in the framework of General Relativity, the Cosmological Principle is discarded to solve the horizon, flatness and cosmological constant problems, without any inflationary assumption, while keeping the successfull predictions of the standard model.
 
\section{The Cosmological Principle}

	The standard justification of this principle rests on two arguments. One, observation-based, is the quasi-isotropy of the CMBR around us. The second, philosophical, asserts that our location in the universe is not special, and what we observe around us must be thus observed the same from everywhere. The standard conclusion is that the matter distribution of the universe is homogeneous. \\

	One can, however, refuse the copernician argument. It is what is done here, where the homogeneity assumption is replaced by another way of taking into account the observed anisotropy of the CMBR: the universe is spherically symmetric and inhomogeneous around us who are located more or less near its center.

\section{An inhomogeneous ``delayed Big-Bang''}

	In an expanding universe, going backward along the parameter called ``cosmic time''$t$ means going to growing energy densities $\rho$ and temperatures $T$. Starting from our present matter dominated age, defined by T $\sim$ 2.73 K, one reaches an epoch where the radiation energy density overcomes the matter one. \\

	To deal with the horizon problem, we have to compute light cones. As will be further shown, this can be done by means of light cones mostly located inside the matter dominated area. In this first approach, the Tolman-Bondi \cite{to}$^{\!,\,}$\cite{bo} solution for spherically symmetrical dust (equation of state $p=0$) models is therefore retained.

\subsection{Class of Tolman-Bondi models retained}

As the observed universe does not present appreciable spatial curvature, it can be approximated by a flat Tolman-Bondi model. \\

	In comoving coordinates ($r, \theta , \varphi$) and proper time $t$, its Bondi line-element is:
\begin{equation}
ds^2 = -c^2 dt^2 + R'^2(r,t)dr^2 + R^2 (r,t)(d\theta^2 + \sin^2 \theta
d \varphi^2)  \label{eq:1}
\end{equation}

	The arbitrary ``mass'' function $M(r)$ can be used to define the radial coordinate $r$:
\begin{equation}
M(r)=M_0 r^3 \qquad  M_0=const. \label{eq:3}
\end{equation}

	Einstein's equations thus give:
\begin{equation}
R(r,t)=\left({9GM_0\over 2}\right)^{1/3} r[t-t_0(r)]^{2/3} \label{eq:4}
\end{equation}

        This model presents two singularity surfaces: \\
$t=t_0(r)$, interpreted as the Big-Bang surface, for which $R(r,t)=0$. \\
$t=t_0(r) + {2\over 3}rt'_0(r)$, usually refered to as ``shell-crossing surface'', for which  \hbox{$R'(r,t)=0$.} \\

	With the choice: $t_0(r=0)=0$, increasing $t$ means going from the past to the future. \\

	From Einstein's equations, the energy density is:
\begin{equation}
\rho (r,t)={1\over 2\pi G[3t-3t_0(r)-2rt'_0(r)][t-t_0(r)]} \label{eq:6}
\end{equation}

\subsection{Singularities}

	Above expression (\ref{eq:6}) implies that the energy density goes to infinity not only on the Big-Bang surface, but also on the shell-crossing surface. A rapid calculation shows that the invariant scalar curvature also takes infinite values on these surfaces. They can thus be both considered as physical singularities. \\

       As will be seen further on, we here need $t'_0(r)>0$. In this case, the shell-crossing surface is situated above the Big-Bang in the $(r,t)$ plane. \\

	For cosmological applications, shell-crossing is not an actual pro\-blem. As energy density increases while reaching the neighbourhood of the shell-crossing surface from higher values of $t$, radiation becomes the dominant component, pressure can no more be neglected, and Tolman-Bondi models no longer hold. Furthermore, in the following, the light cones of interest never leave the \hbox{$t>t_0(r)+{2\over 3}rt'_0(r)$} region.

\subsection{Definition of the temperature}

To provide local thermodynamical equilibrium, the assumption made is that the characteristic scale of the $\rho$ inhomogeneity is much larger than the characteristic length of the photon-baryon interaction. \\

	The local specific entropy is defined as:
\begin{equation}
S(r) \equiv {k_B n_\gamma (r,t)m_b \over \rho (r,t)}   \label{eq:7}
\end{equation}
and the local temperature T by:
\begin{equation}
n_\gamma = a_n T^3 \label{eq:8}
\end{equation}

	One thus obtains, from expression (\ref{eq:6}), an equation for the $T=const.$ surfaces of interest:
\begin{equation}
t=t_0(r)+{r\over 3}t'_0(r)+{1\over 3} \sqrt{r^2t'^2_0(r)+
{3S(r)\over 2\pi G k_B a_n m_b T^3}}   \label{eq:9}
\end{equation}

\subsection{The nearly centered-Earth assumption}

	In this first approach, the Earth is assumed situated sufficiently close to the center of the universe so as to justify the approximation: $r_p=0$ (the subscript ``p'' refering to ``here and now''). \\

	Adding to the specifications of the $t_0(r)$ function:
\begin{eqnarray}
rt'_0 |_{r=0}=0  \nonumber
\end{eqnarray}
one can show, from equation (\ref{eq:9}), that, in the vicinity of $r=0$, the evolution scenario of the universe approximately reproduces the hot Big-Bang one. \\

	The centered-Earth assumption is not an inevitable feature of the model. In a work to be published \cite{sc}, it is  shown that the dipole and quadrupole moments of the CMBR temperature anisotropies can be reproduced with an inhomogeneous Big-Bang function of which the gradient can be chosen all the smaller as the location  of the observer is removed from the center of the universe. \\

        Other work is in progress to show that the horizon problem can be solved with an observer located off this center.

\section{Solving the horizon problem}

	Light travels from the last scattering surface to a present local observer on a light cone going from ($r_p=0$, $t_p$) to a 2-sphere ($r_{es}$, $t_{ls}$) on the last scattering 3-sphere, defined by $T=4000$ K. \\

	To solve the horizon problem, it is sufficient to show that this 2-sphere can be contained inside the future light cone of any ($r=0$, $t>0$)
point of space-time, thus restauring causality before reaching the singularity. \\

	Consider a $t_0(r)$ function, such as the shell-crossing surface is situated above the Big-Bang surface and monotonously increasing with $r$. \\

	The equation for the radial null geodesic is:
\begin{equation}
{dt\over dr}=\pm {1\over 3c} \left(9G M_0\over 2 \right)^{1/3}
{3t-3t_0(r)-2rt'_0(r) \over [t-t_0(r)]^{1/3}}   \label{eq:10}
\end{equation}

	These curves possess an horizontal tangent on and only on the shell-crossing surface where:
\begin{equation}
3t-3t_0(r)-2rt'_0(r)=0  \label{eq:11}
\end{equation}

%\begin{figure}
%\includegraphics[height=15cm,width=11cm,angle=-90]{pgplotc.ps} \\

%\vskip1truecm
%{\bf Fig.2. Solving the horizon problem.} \\
%The dashed lines are the Big-Bang and (upper) shell-crossing surfaces; the plo%tted ones, the constant temperature surfaces: $T=2.73 K$ from point $(r_p=0,t_%p)$ and the last scattering $T=4 000K$ underneath. The solid curves represent %the backward light cones, from $(r_p=0,t_p)$ to the shell-crossing, and from a%ny point P above this surface to the center $(r_c=0, t_c)$.
%\end{figure}

The past light cone $t(r)$ from ($r_p$, $t_p$) verifies equation (\ref{eq:10}) with the minus sign. As it is situated above the Big-Bang and shell-crossing surfaces, $3t-3t_0(r)-2rt'_0(r)$ and $t-t_0(r)$ remain positive. $t(r)$ is thus a strictly decreasing function of $r$ and crosses the shell-crossing surface at a finite point where its derivative goes to zero. On its way, it crosses in turn each $T=const.$ surface at a finite point, which is labelled ($r_{ls1}$, $t_{ls1}$) on the last scattering. \\

	Now consider a backward null radial geodesic from any point P located above the shell-crossing surface, solution of equation (\ref{eq:10}) with the plus sign. Its derivative remains positive as long as it does not reach the shell-crossing surface. Since this surface is strictly increasing with $r$, it cannot be horizontally crossed from upper values of $t$ by a strictly increasing curve. This geodesic reachs therefore $r=0$, without crossing the shell-crossing surface, at $t_c>0$. \\

	This holds for every light cone issued from any point on the last scattering surface. Every point on this sphere can thus be causally connected, and, in particular, the point ($r_{ls1}$, $t_{ls1}$), representing the CMBR as seen from the observer. \\

\section{Examples of appropriate Big-Bang functions}

	The conditions previously imposed upon the Big-Bang function $t_0(r)$ can be summarized as:
\begin{eqnarray}
t_0(r=0) &=& 0  \nonumber \\
t'_0(r) &>& 0  \qquad  \hbox{for all} \qquad r  \nonumber \\
5t'_0(r) + 2rt"_0(r) &>& 0 \qquad \hbox{for all} \qquad r \nonumber \\
rt'_0|_{r=0} &=& 0  \nonumber \\
\nonumber
\end{eqnarray}

	A class of functions fulfilling these conditions is:
\begin{equation}
t_0(r)=br^n  \qquad b>0  \qquad n>0 \label{eq:12}
\end{equation}

        In these models, matter is the dominating component for, at least, 99\% of the cosmic time $t$ elapsed between the last scattering and the shell-crossing surfaces, where are located the pertinent cones. The dust approximation can therefore be considered as correct.\\

%\begin{figure}
%\includegraphics[height=15cm,width=11cm,angle=-90]{pgplot8d.ps}

%\vskip1truecm
%{\bf Fig.3. Case $n=2$, $b=5$x$10^{14}$.} \\
% The solid lines 
%are the light cones, the upper one being  the null geodesic issued from 
%($r_{\rm p},t_{\rm p}$) ending at ($r_{\rm ls1},t_{\rm ls1}$) on the 
%last scattering surface, the lower one being the null cone from 
%($r_{\rm ls1},t_{\rm ls1}$) to ($r=0,t_{\rm c}$). The dashed line 
%represents the Big-Bang  surface, and the plotted one, the last 
%scattering and shell-crossing surfaces which cannot be resolved 
%at the scale of the figure.
%\end{figure}

%\begin{figure}
%\includegraphics[height=15cm,width=11cm,angle=-90]{pgplot9d.ps}

%\vskip1truecm
%{\bf Fig.4. Fig.3 zoomed on small values of $r$ and $t$.} \\
%The solid line is the light cone  from 
%($r_{\rm ls1},t_{\rm ls1}$) to ($r=0,t_{\rm c}$). The dashed line 
%represents the Big-Bang surface, the dotted one, the last scattering 
%surface and the dash-dot-dashed one, the shell-crossing and the 
%radiation-matter equality surfaces which cannot be resolved at the 
%scale of the figure.
%\end {figure}

\section{Conclusion}

	In this preliminary work, it has been shown that the horizon problem can be solved with an inhomogeneous singularity and no inflationary phase. \\

	To deal with the other major problems of standard cosmology, it can be stressed that: \\
- the flatness and cosmological constant problems both proceed from Friedmann's equations and are thus irrelevant in an inhomogeneous model. \\
- the monopoles problem can be considered as a particle physics problem. \\
- perturbations on the Big-Bang surface can produce density fluctuations at all scales which could account for the origin of structure formation. \\

	Observational data at large reshift will have to be analysed in the forma\-lism of inhomogeneous cosmology to discriminate between the standard homogeneous plus inflation paradigm and the new solution here proposed. Such an analysis will yield limits upon the model parameters. \\


\section*{References}
\begin{thebibliography}{99}

\bibitem{cs}M.N. C\'el\'erier and J. Schneider, \Journal{\PLA}{249}{37}{1998}.

\bibitem{to}R.C. Tolman, \Journal{\it Proc. Nat. Acad. Sci. USA}{20}{1342}{1934}.

\bibitem{bo}H. Bondi, \Journal{\it Month. Not. Roy. Astr. Soc.}{107}{410}{1947}.

\bibitem{sc}J. Schneider and M.N. Célérier, {\it Astron. Astrophys.} in press, (1999).


\end{thebibliography}
\end{document}